\documentclass[fleqn,10pt]{wlscirep}
\usepackage{lineno}

\usepackage{mathrsfs}
\usepackage{amssymb, amsbsy, amsmath, latexsym, dsfont, array, layout, mathrsfs, color}

\usepackage{mathtools}

\usepackage{amsmath,amssymb}
\usepackage{bm}
\usepackage{graphicx}
\usepackage{braket,dsfont}
\usepackage{color}
\usepackage[10pt]{moresize}
\usepackage{mathrsfs}
\usepackage{multirow}
\usepackage{hyperref}
\usepackage{changes}
\usepackage{mathdots}
\let\oldcaption\caption
\renewcommand{\caption}{\sffamily \oldcaption}

\newcounter{lastnote}

\begin{document}

\title{Experimental Quantum Network Coding}
\author[1,2,3]{He Lu}
\author[1,2]{Zheng-Da Li}
\author[1,2]{Xu-Fei Yin}
\author[1,2]{Rui Zhang}
\author[3]{Xiao-Xu Fang}
\author[1,2]{Li Li}
\author[1,2]{Nai-Le Liu}
\author[1,2,*]{Feihu Xu}
\author[1,2,*]{Yu-Ao Chen}
\author[1,2,*]{Jian-Wei Pan}

% Place the author information here.  Please hand-code the contact
% information and notecalls; do *not* use \footnote commands.  Let the
% author contact information appear immediately below the author names
% as shown.  We would also prefer that you don't change the type-size
% settings shown here.

\affil[1]{Shanghai Branch, National Laboratory for Physical Sciences at Microscale and Department of Modern Physics, University of Science and Technology of China, Shanghai 201315, China}
\affil[2]{Synergetic Innovation Center of Quantum Information and Quantum Physics, University of Science and Technology of China, Hefei, Anhui 230026, China}
\affil[3]{School of Physics, Shandong University, Jinan 250100, China}
\affil[*]{Correspondence: Feihu Xu (feihuxu@ustc.edu.cn); Yu-Ao Chen (yuaochen@ustc.edu.cn); Jian-Wei Pan (pan@ustc.edu.cn)}

%%%%%%%%%%%%%%%%% END OF PREAMBLE %%%%%%%%%%%%%%%%
%Double-space the manuscript.

%\baselineskip24pt
\renewcommand{\baselinestretch}{1.2}


\begin{abstract}
%\internallinenumbers
Distributing quantum state and entanglement between distant nodes is a crucial task in distributed quantum information processing on large-scale quantum networks. Quantum network coding provides an alternative solution for quantum state distribution especially when the bottleneck problems  must be considered and high communication speed is required.  Here, we report the first experimental realization of quantum network coding on  the butterfly network. With the help of prior entanglements shared between senders, two quantum states can be  transmitted perfectly through  the butterfly network. We {demonstrate this protocol by employing eight photons generated} via spontaneous parametric down-conversion. We observe cross-transmission of {single-photon} states with an average fidelity of $0.9685\pm0.0013$,  and that of two-photon entanglement with an average fidelity of $0.9611\pm0.0061$, both of which are  greater than the theoretical upper bounds without prior entanglement.
\end{abstract}

\maketitle

%\begin{linenumbers}
\section*{Introduction}
The global quantum network~\cite{Kimble2008} is believed to be the next-generation information processing platform  and promises an exponential increase in computation speed, a secure means of communication~\cite{Lo2014,Feihu19} and an exponential saving in transmitted information~\cite{Buhrman2001}. The efficient distribution of quantum state and entanglement is a key ingredient for such a global platform.  Entanglement distribution\cite{Pan98} and quantum teleportation\cite{Bennett1993} can be employed to transmit quantum states over long  distances.  By exploiting entanglement swapping\cite{Pan98} and quantum purification, the  transmission distance could be extended  significantly and the fidelities of transmitted states can be enhanced up to unity, which is  known as quantum repeaters\cite{Dur99}. However, with the  increased of complexity of quantum  networks, especially when many parties  require simultaneous communication and communication rates exceed the capacity of quantum channels, low transmission rates or long delays, known as bottleneck problems,  are expected to occur.Thus, it is important to resolve the bottleneck problem and achieve high-speed quantum communication. This question is in the line with  issues related to quantum communication complexity, which  attempts to reduce the amount of information  to be transmitted to solve distributed computational tasks~\cite{Buhrman2010}.

The bottleneck problem is common in classical networks. A landmark solution is the network coding  concept~\cite{ACLY},  where the key idea is to allow coding and replication of information locally at any intermediate node  in the network. The metadata arriving from two or more sources at intermediate nodes  can be combined into a single packet,  and this distribution method can increase the effective capacity of a network by minimizing the number and severity of bottlenecks. The improvement is most pronounced when  the network traffic volume is near the maximum capacity obtainable  via traditional routing. As a result, network coding  has realized a new communication-efficient  method to send information through networks~\cite{Ho2008}.

A primary question  relative to quantum  networks is whether network coding is possible for quantum state transmission, which is referred as quantum network coding (QNC).  Classical network coding  cannot be applied  directly in a quantum case  due to the no-cloning theorem\cite{Wootters1982}. However, remarkable theoretical effort has been  directed at this important question.  For example, Hayashi \emph{et al.} \cite{HINRY}  were the first to study QNC, and   they proved that perfect quantum state cross-transmission is impossible in the butterfly network, i.e., the fidelity of crossly transmitted quantum states cannot reach one. However, if two senders  have share entanglements  priorly, the perfect QNC is possible  by exploiting quantum teleportation~\cite{LOW,Hayashi2007,Kobayashi2011}.  Thus, various studies have focused on network coding for quantum networks, such as the multicast problem~\cite{Shi,Kobayashi2010}, QNC based on quantum repeaters~\cite{Satoh2012},  QNC based quantum computation\cite{Soeda2011} and other efficient quantum-communication protocols with entanglement~\cite{ACL2007,Epping2016,Schoute2016,Hahn2018}. Despite these theoretical advances, to the best of our knowledge,  an experimental demonstration of QNC has not been realized in  a

\begin{figure}[t!]
\scalebox{1}{\includegraphics[width=\columnwidth]{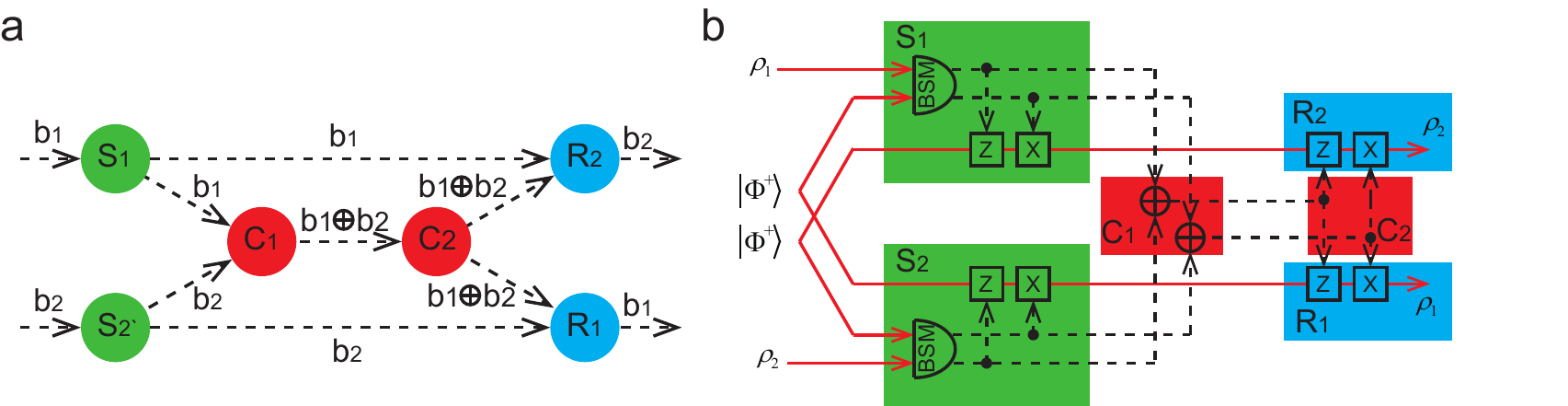}}
%\internallinenumbers
\caption{\textbf{Classical network coding and quantum network coding on a butterfly network. a,} classical network coding on a butterfly network. Dash line with arrow represents information flow with a capacity of a single packet. In the two simultaneous unicast connections problem, one packet $b_1$ presented at source node $S_1$ is required to  transmit to node $R_1$ and the other packet $b_2$ presented at source node $S_2$ is required to  transmit to node $R_2$ simultaneously. The intermediate node $C_1$ performs a coding operation XOR $\oplus$ on $b_1$ and $b_2$. $C_2$ makes copies of $b_1\oplus b_2$ and sends them to $R_1$ and $R_2$ respectively. $R_1$ and $R_2$ decode by performing further XOR operations on the packets that they each receive. \textbf{b, }quantum network coding on butterfly network. The red line with arrow represents quantum information flow with a capacity of a single qubit, and the dash line with arrow represents classical information flow with a capacity of a two bits. See main text for more details.} \label{fig:concept}
\end{figure}

\noindent laboratory, even for the simplest  of cases.

In this  study, we  provide the first experimental demonstration of a perfect QNC protocol on the butterfly network.  This experiment  adopted the protocol proposed by Hayashi~\cite{Hayashi2007}, who proved that perfect QNC is achievable if the two senders have two prior maximally-entangled pairs, while it is impossible without prior entanglement. We demonstrate this protocol by employing eight photons generated via spontaneous parametric down-conversion (SPDC). We observed a cross-transmission of single-photon states with  an average fidelity of $0.9685\pm0.0013$,  as well as cross-transmission of two-photon entanglement with  an average fidelity of $0.9611\pm0.0061$, both of which are  greater than the theoretical upper bounds without prior entanglement.

% \end{linenumbers}

% entanglement fidelity = 0.9256; state transmission fidelity = 0.9504.

%Hayashi proposed a protocol which enables perfect quantum transmission with prior entanglement between two senders only, while it is impossible without prior entanglement~\cite{Hayashi2007}.

% \begin{linenumbers}

% The receivers decode by performing further coding operations on the packets that they each receive.

\section*{Results}
\subsection*{QNC on butterfly network}

Network coding refers to coding at a node in a network~\cite{ACLY}. The most famous example of network coding is the butterfly network,  which is illustrated in Fig.~\ref{fig:concept}a. While network coding has been generally considered for multicast in a network, its throughput  advantages are not limited to multicast. We focus on a simple modification of the butterfly network  that facilitates an example  involving two simultaneous unicast connections. This is also known as \emph{2-pairs problem}~\cite{Yeung1999,Li2004}:  which seeks to answer the following: for  two sender-receiver pairs ($S_1$-$R_1$ and $S_2$-$R_2$), is there a way  to send two messages between the  two pairs simultaneously? In the network  shown in Fig.~\ref{fig:concept}a,  each arc represents a directed link that  can carry a single packet reliably.  Here, is  a single packet $b_1$ present at sender $S_1$ that we  want to  transmit to receiver $R_1$ and a single packet $b_2$ present at sender $S_2$ that we  want to  transmit to receiver $R_2$ simultaneously. The intermediate node $C_1$ breaks from the traditional routing paradigm of packet networks, where intermediate nodes are only  permitted to make copies of received packets for output.  Intermediate node $C_1$ performs a coding operation  that takes two received packets, forms a new packet by taking the bitwise sum or XOR), of the two packets, and outputs the resulting packet $b_1\oplus b_2$.  Ultimately, $R_1$ recovers $b_2$ by taking the XOR of $b_1$ and $b_1\oplus b_2$ and  similarly $R_2$ recovers $b_1$ by taking the XOR of $b_2$ and $b_1\oplus b_2$. Therefore, two unicast connections can be established  with coding and cannot without coding.

In the case of quantum $2$-pairs problem, the model is the same butterfly network (Fig.~\ref{fig:concept}a) with unit-capacity quantum channels and the goal is to send two unknown \emph{qubits} crossly,  i.e., to send $\rho_1$ from $S_1$ to $R_1$ and $\rho_2$ from $S_2$ to $R_2$ simultaneously. However, two rules prevent applying classical network coding directly  in the quantum case: (i)  an XOR operation for two quantum states is not possible; (ii)  an unknown quantum state cannot be cloned exactly. Therefore, it  has been proven that the quantum $2$-pairs problem is impossible~\cite{HINRY}.

% it is impossible to achieve the perfect transmission of two quantum states crossly.

Hayashi  proposed a protocol  that addresses the quantum $2$-pairs problem  by exploition prior entanglements between two senders~\cite{Hayashi2007}. As shown in Fig.~\ref{fig:concept}b, the scheme is a resource-efficient protocol that  only requires two pre-shared pairs of maximally entangled state $\ket{\Phi^{+}}$ between the two senders.  Notice that if the sender ($S_1$, $S_2$) nodes and receiver ($R_1$, $R_2$) nodes allow to

\begin{figure*}[h]
\scalebox{1}{\includegraphics[width=1\columnwidth]{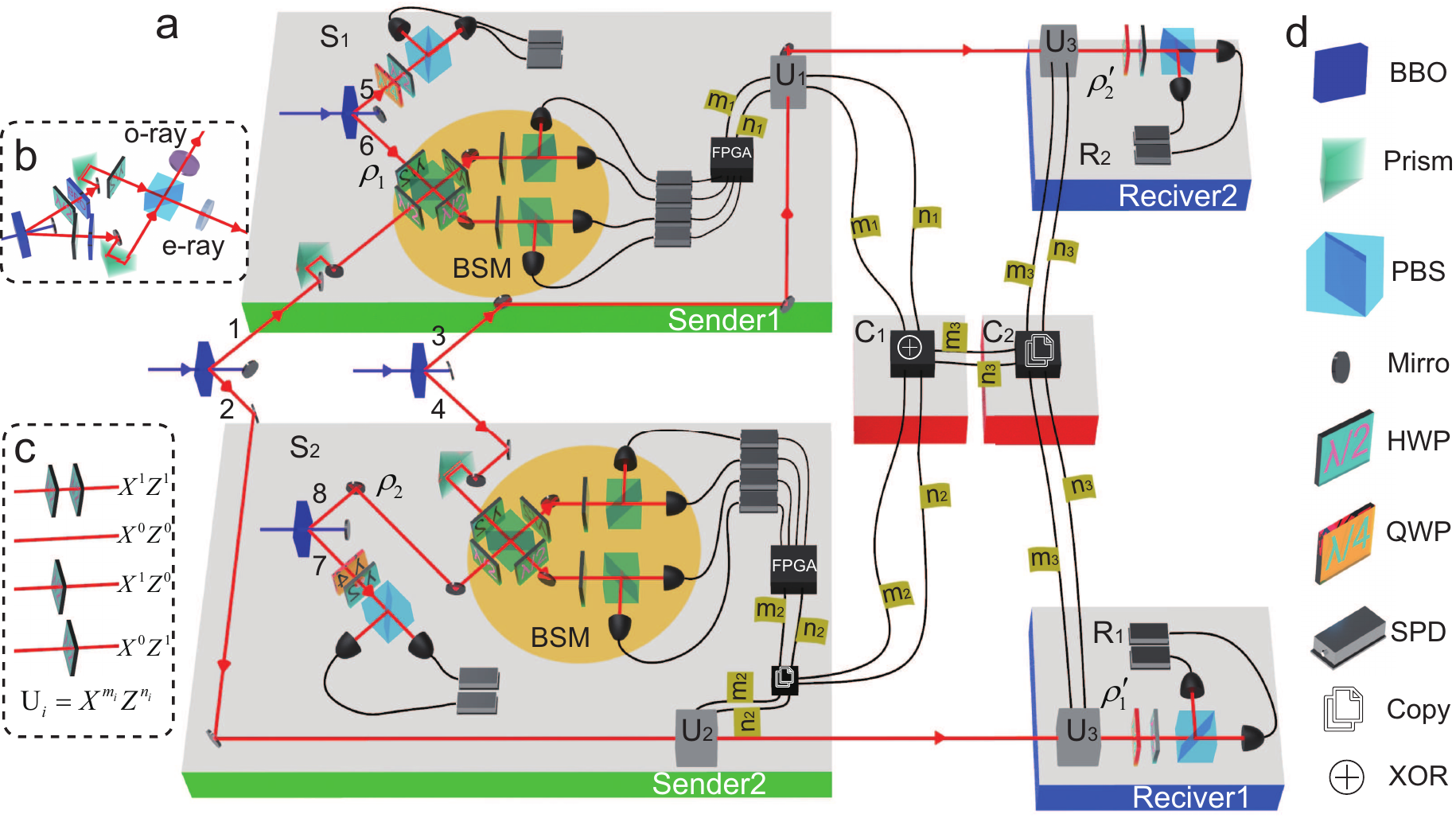}}
%\internallinenumbers
\caption{\textbf{Schematic drawing of the experimental setup. a,} an ultraviolet pulse successively pass through four BBO crystals, and generate four pairs of maximally entangled photons. We use four  Bell-state synthesizer (shown in Fig.\ref{fig:setup}b). to improve the counter rate of entangled photon pair. To avoiding a mess of illustration, we separated propagation of ultraviolet pulse. In our experiment, the ultraviolet pulse is guided by mirrors to shine on four BBO one by one.  All the photons are collected by single-mode fiber and detected by single photon detecters (SPD). The coincidence is recored by several home-made field-programmable gate arrays (FPGA). See main text for more details.  \textbf{b,} Bell-state synthesizer. The generated photons are compensated by a HWP at 45$^{\circ}$ and 1-mm-long BBO crystal. Then, one photon is rotated by a HWP at 45$^{\circ}$ and finally two photons are recombined on a PBS. With Bell-state synthesizer makes ordinary ray(o-ray) exiting from one port of PBS and extraordinary ray(e-ray) exiting the other port of PBS. \textbf{c,} the unitary operation $U_i=X^{m_i}Z^{n_i}$ is realized by HWPs. We post-selectively apply $U_i$ according to $m_in_i$. \textbf{c, } symbols used in \textbf{a}, \textbf{b} and \textbf{c}. BBO: Beta barium borate crystal. PBS: Polarizing beam splitter. HWP: Half-wave plate. QWP: Quarter-wave plate. SPD: Single photon detector.}
 \label{fig:setup}
\end{figure*}

\noindent share prior entanglements, then transmitting classical information with classical network coding can complete the task only by using quantum teleportation\cite{LOW}. If free classical between all nodes is not limited, perfect 2-pair communication over  the butterfly network is possible\cite{Kobayashi09}. However, we consider a more practical situation that the sender and receiver nodes do \emph{not} share any prior entanglements. Also, the channel capacity is limited to transmit either one qubit or two classical bits. Hayashi proved that the average fidelity of quantum state transmitted is upper bounded by 0.9504  for single-qubit state and 0.9256 for entanglement without prior entanglement\cite{Hayashi2007}. However, with prior entanglement between senders, the average fidelity can reach 1. The protocol is summarized  as follows (see Fig.~\ref{fig:concept}b).

\begin{enumerate}
  \item $S_1$ ($S_2$) applies the Bell-state measurement (BSM) between the transmitted state $\rho_1$ ($\rho_2$) and one qubit of $\ket{\Phi^{+}}$. According to the result of BSM $m_1n_1$ ($m_2n_2$), $S_1$ ($S_2$) perform  the unitary operation $X^{m_1}Z^{n_1}$($X^{m_2}Z^{n_2}$) on the other qubit of $\ket{\Phi^{+}}$ .

  \item $S_1$ ($S_2$) sends the quantum state (after  the unitary operation) to $R_2$ ($R_1$), and sends the classical bits $m_1n_1$ ($m_2n_2$) to $C_1$. $C_1$ performs the XOR on $m_1$ and $m_2$, $n_1$ and $n_2$ respectively, then sends $m_3=m_1\oplus m_2$ and $n_3=n_1\oplus n_2$ to $C_2$. $C_2$  makes copies of $m_3n_3$ and sends  them to $R_1$ and $R_2$, respectively.

  \item $R_1$ and $R_2$  recover the quantum states $\rho_1$ and $\rho_2$ by applying  the unitary operation $X^{m_3}Z^{n_3}$ on their received quantum states.
\end{enumerate}

\begin{figure*}[t!]
\scalebox{1}{\includegraphics[width=\columnwidth]{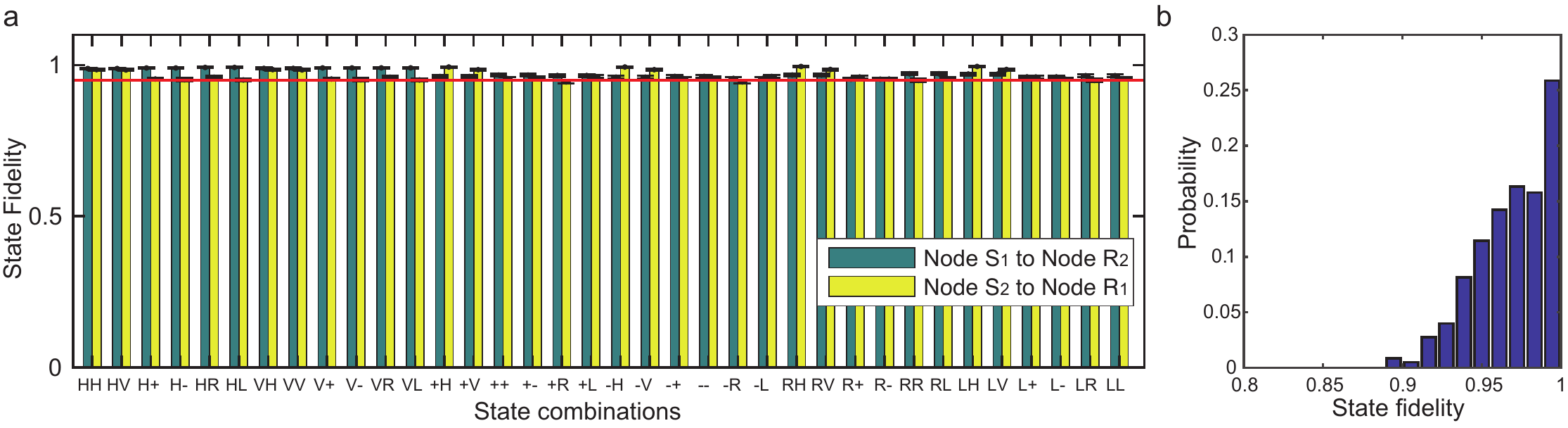}}
%\internallinenumbers
\caption{\textbf{Fidelities of crossly transmitted quantum states. a,}  The green bar represents $F_{S_1\to R_2}$, and the yellow bars represents $F_{S_2\to R_1}$. The pair-appeared bars  represent fidelities measured  simultaneously at $R_1$ and $R_2$. For example, $HR$ means that $S_1$ delivers $\ket{H}$ and $S_2$ delivers $\ket{R}$. The red line represents the threshold of $F_ th=0.9503$.  The fourfold coincidence is  approximately 1.5 counts per second. We accumulate coincidences for 240 seconds and  a total  of 720 counts for each two-state transition. The error bars are calculated assuming a Poisson statistics for the coincidence counts and Gaussian error propagation. \textbf{b,}  Histogram of state fidelities.} \label{fig:fidelity1}
\end{figure*}

\subsection*{Experimental realization}
We demonstrate the perfect QNC protocol by employing the polarization degree of freedom of photons generated via SPDC. As shown in Fig.~\ref{fig:setup}a, an ultraviolet pulse (with  a central wavelength of 390 nm,  power of 100 mW, pulse duration of 130 fs and repetition of 80 MHz) successively passes through four  2-mm-long BBO crystals  successively and generates four maximally entangled photon pairs via SPDC in the form of  $\ket{\Psi^{+}}_{ij}=1\sqrt{2}(\ket{HV}+\ket{VH})_{ij}$. Here $H$ ($V$) denotes the horizontal (vertical) polarization and $i$, $j$ denote the path modes. Then, we use a Bell-state synthesizer to reduce the frequency correlation between two photons\cite{Yoonho03, Yao2012} (as shown in Fig.~\ref{fig:setup}b). After  the Bell-state synthesizer,  $\ket{\Psi^{+}}_{ij}$ is converted to $\ket{\Phi^{+}}_{ij}=1\sqrt{2}(\ket{HH}+\ket{VV})_{ij}$. We set narrow-band filters with full-width at half maximum ($\lambda_{FWHM}$) of 2.8 nm and 3.6 nm for the e- and o-ray, respectively,  and, with this filter setting, we observe an average two-photon coincidence count rate of 21000 per second with a visibility of 99.6\% in the $\ket{H(V)}$ basis and  visibility of 99.0\% in the $\ket{+(-)}$ basis, from which we calculate the fidelity of prepared entangled photons with  an ideal $\ket{\Phi^{+}}$ of 99.3\%. We estimate a single-pair generation rate of $p\approx 0.0036$ and overall collection efficiency of 28\%.

$\ket{\Phi^{+}}_{12}$ and $\ket{\Phi^{+}}_{34}$ are the two entangled pairs  priorly shared between $S_1$ and $S_2$ , i.e., $S_1$ holds photon 1\&3 and  $S_2$ holds photon 2\&4. $\ket{\Phi^{+}}_{56}$ and $\ket{\Phi^{+}}_{78}$ are held by $S_1$ and $S_2$, respectively. Photon 5 is projected on $\alpha_1^{*}\ket{H}+\beta_1^{*}\ket{V}$ to prepare $\rho_{1}$ with ideal form in $\alpha_1\ket{H}+\beta_1\ket{V}$. Similarly,  photon 7 is projected on $\alpha_2^{*}\ket{H}+\beta_2^{*}\ket{V}$ to prepare $\rho_{2}$ with ideal form $\alpha_2\ket{H}+\beta_2\ket{V}$.

On $S_1$'s side, by finely adjusting the position of  the prism on the path of photon 1, we interfere  with photons 1 and photon 6 on a polarizing beam splitter (PBS) to realize a Bell-state measurement (BSM). The BSM projects  photons 1 and photon 6 to $\ket{\psi}\in\{\ket{\Psi^{+}}, \ket{\Psi^{-}}, \ket{\Phi^{+}}, \ket{\Phi^{-}}\}$.  As the complete BSM is impossible with linear optics, we  perform the complete measurements with two setup settings by  rotating the angle of  the half-wave plate (HWP) on path 6 or 1 before they interfere from 0$^{\circ}$ to 45$^{\circ}$.  Note that in each setup, the success probability to identify two of the Bell states is 50\%. So, the total success probability is 25\% in our experiment. The BSM  results (different responses on the four detectors after the interference)  are related to two classical bits denoted as $m_1n_1\in\{00, 01, 10, 11\}$. According to the BSM results, $S_1$ applies the unitary operation $U_1=X^{m_1}Z^{n_1}$ on photon 3, and then sends $m_1n_1$ to node $C_1$ and photon 3 to the receiver node $R_2$.  Here, we use $X$, $Y$, $Z$ to represent the Pauli-$X$, Pauli-$Y$, Pauli-$Z$ matrix. Similarly, on  the $S_2$ side, we interfere  with photons 4 and 8 on a PBS to realize a BSM with result of $m_2n_2$, according to which $S_2$  applies the unitary operation $U_2=X^{m_2}Z^{n_2}$ on photon 2. Then, $S_2$ sends $m_2n_2$ to node $C_1$ and  sends photon 2 to the receiver node $R_1$.

On node $C_1$, we  perform the XOR operation on $m_1$ and $m_2$ and $n_1$ and $n_2$, and send the results $m_3=m_1\oplus m_2$, $n_3=n_1\oplus n_2$ to node $C_2$, where we make two copies of $m_3n_3$ and  send these copies to $R_1$ and $R_2$. Finally, according to $m_3n_3$, we apply  the unitary operation $U_3=X^{m_3}Z^{n_3}$ on  photons 3 and photon 2 to recover $\rho_{2}$ and $\rho_{1}$.

In our experiment, the unitary operation is realized by HWPs with transformation matrix $U(\theta)=\begin{pmatrix}
cos 2\theta & sin 2\theta \\
sin 2\theta & -cos 2\theta \\
\end{pmatrix}$, where $\theta$ is the angle fast axis  relative to  the vertical axis. $X^0Z^0=I$ means no operation on  the photon.  Here, $X^0Z^1=Z$ is realized by setting  an HWP at $0^{\circ}$. $X^1Z^0=X$ is realized by setting  an HWP at $45^{\circ}$,  and $X^1Z^1=XZ$ is realized by setting two HWPs (one at $45^{\circ}$ and the other at $0^{\circ}$ (Shown in Fig.\ref{fig:setup}c))

\begin{figure*}
\centering
\scalebox{1}{\includegraphics[width=0.9\columnwidth]{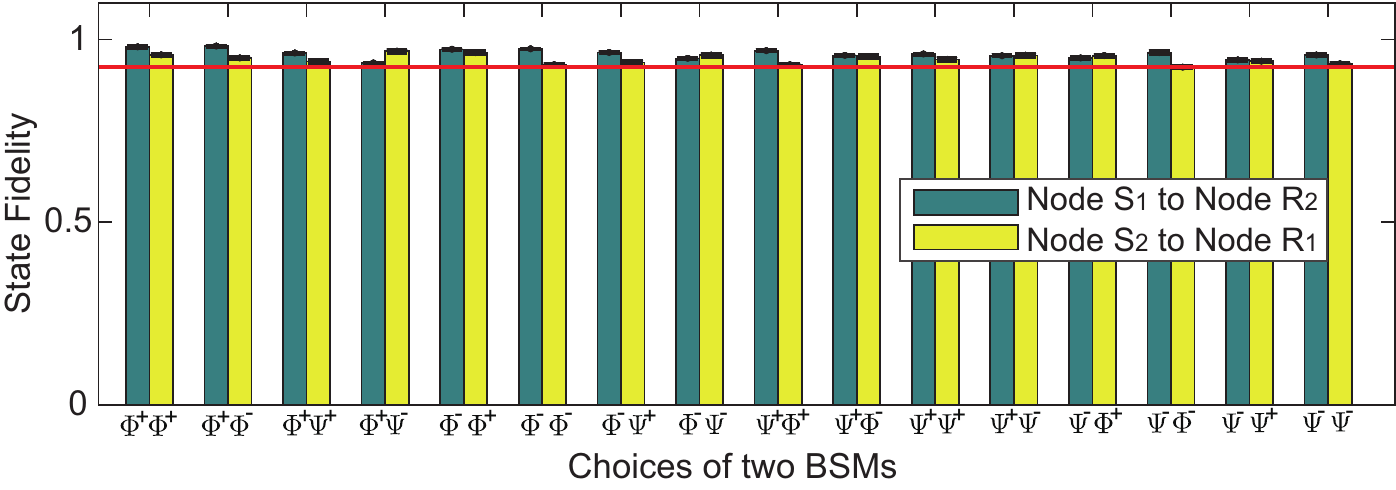}}
%\internallinenumbers
\caption{\textbf{Fidelities of crossly established entanglement.} The green bar represents entanglement established between $S_1$ to $R_2$, and the yellow bars represents $S_2$ to $R_1$. The pair-appeared bars represents the fidelities measured simultaneously. For each BSM, there are four possible outcomes, thus there are 16 situations. The red line represents the threshold of $F_ {th}=0.9256$.  The fourfold coincidence is about 3 counts per second. We accumulate coincidences for 120 seconds and total 720 counts for each two-state transition. The error bars are calculated assuming a Poisson statistics for the coincidence counts and Gaussian error propagation. } \label{fig:fidelity2}
\end{figure*}

\subsection*{Experimental results}
We first show that two single-photon states can be crossly delivered from $S_1$ to $R_2$ and from $S_2$ to $R_1$ simultaneously in the butterfly network. $S_1$ and $S_2$  can prepare six individual quantum states  $\rho_1$ and $\rho_2$ with  an average fidelity of 99.3\%. $\rho_1$ and $\rho_2$  have an ideal form of $\rho_1=\ket{\phi_1}\bra{\phi_1}$ and $\rho_2=\ket{\phi_2}\bra{\phi_2}$, where $\ket{\phi_1}, \ket{\phi_2}\in\{\ket{H}, \ket{V}, \ket{\pm}=\frac{1}{\sqrt{2}}(\ket{H}\pm\ket{V}), \ket{L(R)}=\frac{1}{\sqrt{2}}(\ket{H}\pm i\ket{V})\}$. In our experiment, both $S_1$ and $S_2$  irrelatively  select  $\rho_1$ and $\rho_2$ from six states for transmission, thereby resulting in a total of 36 combinations. After  recover of $R_1$ and $R_2$, we measure the fidelities between the recovered state $\rho^{\prime}_{1}$ ($\rho^{\prime}_{2}$) and  the ideal input state $\rho_{1}=\ket{\phi_1}\bra{\phi_1}$ ($\rho_{2}=\ket{\phi_2}\bra{\phi_2}$), i.e., $F_{S_1\to R_2}=Tr(\ket{\phi_1}\bra{\phi_1}\rho^{\prime}_{1})$ and $F_{S_2\to R_1}=Tr(\ket{\phi_2}\bra{\phi_2}\rho^{\prime}_{2})$.  We project  the photon on  the $\ket{\phi}(\ket{\phi_ \perp})$ basis and  record the counts $N_+$ and $N_-$, where $\ket{\phi_ \perp}$ is the orthogonal state of $\ket{\phi}$. Thus, the fidelity of  the transferred single-photon state  can be calculated by $F=\frac{N_+}{N_++N_-}$. The average fidelities over all possible  BSM outcomes are shown in Fig.~\ref{fig:fidelity1}a. Note that each BSM has four possible outcomes, thus there are 16 combinations of outcomes for the two BSMs. For each combination, we apply the unitary operations and record the measured fidelities.  In Fig.~\ref{fig:fidelity1}a, the red line represents the theoretical upper bound of the average fidelity without prior entanglement, i.e., $F_{th}=0.9503$.  Specifically, Fig.~\ref{fig:fidelity1}b shows the histogram of all measured fidelities of the 576 situations, and the average fidelity is quantified as $\bar{F}=\sum\limits_ip_iF_i=0.9685\pm0.0013$, where $p_i$ and $F_i$ are the probability and fidelity shown in Fig.~\ref{fig:fidelity1}b.  The average fidelity beyonds $F_{th}=0.9503$ with 14 standard deviations.

 We also show that two-photon entanglement  can be established crossly with  this setup, i.e., two-photon entanglement  can be established between  $S_1$ and $R_2$ and $S_2$ and $R_1$ simultaneously.  Here, the experimental setup is the same, $S_1$($S_2$) does not project photon 5(7) on $\alpha^{\ast}\ket{H}+\beta^{\ast}\ket{V}$,  and photon 5(7) is retained to perform the joint measurements with photon 2(3). To quantify the cross entanglement between photon 5  and 2 and 7  and 3, we measure the entanglement witness on $rho_{52}$ and $\rho_{73}$, respectively. In particular, we measure the entanglement witness $ W=I/2-\ket{\Phi^{+}}\bra{\Phi^{+}}$, which can also be related to the entanglement fidelity $\langle W\rangle=1/2-F_{ent}$. Here $F_{ent}$ is defined as the entanglement fidelity between the entanglement state $\rho_{ij}$ and the maximal entanglement state $\ket{\Phi^{+}}$, i.e., $F_{ent}=Tr(\rho_{ij}\ket{\Phi^{+}}\bra{\Phi^{+}})$. $\ket{\Phi^{+}}\bra{\Phi^{+}}$ can be decomposed to  a local observable as $\ket{\Phi^{+}}\bra{\Phi^{+}}=\frac{II+XX-YY+ZZ}{4}$.  By measuring the expected values of local observables, we can calculate the entanglement fidelity. The local observable $\mathcal O$ can be  expressed as $\mathcal O=\ket{\phi}\bra{\phi}-\ket{\phi_\perp}\bra{\phi_\perp}$, where $\ket{\phi} (\ket{\phi_\perp})$ is the eigenstate of $\mathcal O$ with eigenvalue of 1(-1). The expected value of $\mathcal O$  can be calculated by the counts $\langle\mathcal O\rangle=\frac{N_{+}-N_{-}}{N_{+}+N_{-}}$. The experimental results of the fidelities of cross entanglement are shown in Fig.~\ref{fig:fidelity2}. We calculate that the average fidelity of two crossly established is $0.9611\pm0.0061$, which  beyonds 0.9256 with 5.8 standard deviations.
\section*{Discussion}

QNC provides an alternative solution for  the transition of quantum states in quantum networks. Compared  to entanglement swapping, QNC  demonstrates superiority especially when quantum resources are limited or  a high communication rate is required\cite{Satoh2016}.  In addition, large-scale QNC  demonstrates superiority  relative to fidelity-performance as well\cite{Nguyen2017}.  In this paper, We have demonstrated the first perfect QNC on a butterfly network. The average fidelities of cross state transmission and cross entanglement distribution achieved in our experiment  exceed the  theoretical upper bounds  permitted without prior entanglement.  We expect that our results will pave the way  to experimentally  explore the advanced features of prior entanglement in quantum communication.  In addition, we expect that our results will realize opportunities for various studies of efficient quantum communication protocols in quantum networks with complex topologies.

\section*{Author contributions}
H.L., F.X., Y.-A.C. and J.-W.P.  established the theory and  designed the experimental setup. H.L., Z.-D.L., Y.-X.F. and R.Z.  performed the experiment. H.L. X.-X. F., L.L. and N.-L. L.  analyzed the data. H.L., F.X. and Y.-A.C.  wrote the paper with contributions from all authors.

\section*{Competing interests}
The authors declare no competing interests.

\section*{Data availability}
The data  are available from the corresponding author  upon reasonable request.

\section*{Acknowledgement}
This work was supported by the National Key Research and Development (R\&D) Plan of China (grants 2018YFB0504300 and 2018YFA0306501), the National Natural Science Foundation of China (grants 11425417, 61771443), the Anhui Initiative in Quantum Information Technologies and the Chinese Academy of Sciences. H. Lu was partially supported by Major Program of  Shandong Province Natural Science Foundation (grants ZR2018ZB0649). F. Xu thanks Prof. Bin Li for the early inspiration to the subject.

%\bibliography{QNC}

%\end{linenumbers}
\end{document}